\documentstyle [12pt,epsf] {article} 
    
   \catcode`@=12 
\begin{document} 
   \vspace{0.5in} 
   \oddsidemargin -.375in 
   \newcount\sectionnumber 
   \sectionnumber=0 
   \def\be{\begin{equation}} 
   \def\ee{\end{equation}} 
   \def\ber{\begin{eqnarray}} 
   \def\eer{\end{eqnarray}} 
\def\vn{\vec{n}} 
   \thispagestyle{empty} 
   \begin{flushright} 
UT-PT-99-18\\ October 1999\\
   \end{flushright} 
   \vspace {.5in} 
   \begin{center} 
   {\Large \bf Vacuum Stability Higgs Mass Bound Revisited\\ with
Implications for  Extra Dimension Theories \\} \vspace{.5in} 
   {\rm A. Datta$^1$ and X. Zhang$^2$ \\} 
   \vspace{.5in} 
   {$^1$ \it Department of Physics, University of Toronto, \\ 
   Toronto, Ontario, Canada, M56 1A7\\} 

   {$^2$ \it CCAST (World Laboratory), P.O. Box 8730, Beijing 100080 and\\ 
   Institute of High Energy Physics, Beijing 100039, P.R. China\\} 
   \vspace{.1in} 
   \vskip .5in 
   \end{center} 
   \begin{abstract}

We take the standard model to be an effective theory including
higher dimensional operators suppressed by
scale $\Lambda$ and
re-examine the higgs mass bounds
from the requirements of 
vacuum stability.
Our results show that the effects of the higher dimensional
 operators on the higgs mass limits are significant. As an implication 
of our results, we study the
vacuum stability higgs mass bounds in theories with extra dimensions.

   \end {abstract} 
   \newpage 
   \baselineskip 24pt


   One of the important issues in particle physics is to 
   understand the origin of the electroweak scale. In the standard model,
   the 
   electroweak symmetry breaking arises from a complex 
   fundamental higgs scalar. However, the theoretical arguments 
   of "triviality" \cite{1} and "naturalness" \cite {2}, 
   suggest that such a simple spontaneous 
   symmetry breaking mechanism may 
   not be the whole story. This leads to the belief 
   that the higgs sector of the 
   standard model is an effective theory valid below 
   some cut-off scale $\Lambda$. 
   Direct searches  at LEP  has put a
lower bound on the higgs mass $m_h \sim 95.5$ GeV \cite{G} and 
from a global fit to
electroweak precision 
   observables one can put an upper bound on the higgs mass, $m_h <$ 250 GeV
   at 95 \% C.L \cite{Lep}. 
   This assumes that the scale of new physics, $ \Lambda$, 
   is high enough and thus new physics does not have 
significant effects on the electroweak 
   precision observables. 
Recent studies show that this upper bound on the higgs mass 
   may be relaxed if 
   the scale $\Lambda$, which suppresses the higher 
   dimensional operators arising from new physics, is around a few 
TeV \cite {Hall,Barb,Ravi}. There are also
theoretical arguments of triviality and vacuum stability which
place bounds on the higgs mass.
 A lower bound on the higgs mass $\sim 135$ GeV  is
   obtained 
   by requiring the standard model vacuum to be stable to the Planck
scale \cite{Casas}.
  It was shown in Ref
   \cite{DYZ} 
   that, in the presence of higher dimensional operators, the vacuum
stability limit on the higgs mass can also be changed.
    
There are many proposals
   for beyond the standard model physics. They have to address 
   the hierarchy problem which is the understanding of why the weak scale 
   $M_W \sim$ 100 GeV is so much smaller than $M_{Pl} \approx 10^{19}$ GeV, 
   which is believed to be the fundamental mass scale of gravity. An
   exciting 
   solution to the 
   hierarchy problem in recent years is provided by the  
   theories with extra dimensions \cite{Nima}. In such theories 
   space time is enlarged to contain large extra compact spatial dimensions.
   The possibility of lowering the compactification scale $ \sim $TeV
   was discussed in Ref\cite{Anton}. 
   At distances smaller than the size of these extra dimensions the 
   gravitational force varies more rapidly than the inverse square law, so 
   that the fundamental mass scale of gravity, $\Lambda$, can be made much
   smaller 
   than $M_{Pl}$. The hierarchy problem 
   is avoided if this fundamental mass scale is of the order of the weak 
   scale. 

   Below the scale $\Lambda$, the physics can be described by an 
   effective theory where the leading 
   terms in the lagrangian are given by the 
   standard model. The corrections which come from the underlying 
   theory with extra dimensions are described by higher dimension 
   operators, 
   \begin{eqnarray} 
   {\cal L}^{new} = \sum_{i} \frac{c_i}{\Lambda^{d_i-4}} {\cal O}^i , \ 
   \end{eqnarray} 
   where $d_i$ are the dimensions of ${\cal O}^i$, which are 
   integers greater than 4. The operators ${\cal O}^i$ are 
   $SU(3)_c \times SU(2)_L \times U(1)_Y$ invariant and 
   contain only 
   the standard model fields. The dimensionless 
   parameters $c_i$, 
   determining the strength of the contribution of operators ${\cal O}^i$, 
   are expected to be of O(1) or larger \cite{Dine}. 

    In this work, we first present a complete analysis of the higgs mass 
   bound from 
   consideration of vacuum stability using the effective lagrangian approach. 
   We improve upon the analysis of Ref\cite{DYZ} by extending the range
of the
    coefficients $c_i$ from $c_i=(-1) - 0.1$ to the range $|c_i|=1-10$
as expected in theories with large extra dimensions.
We also extend the 
   scale of 
   new physics $\Lambda$ from 20 TeV to $M_{Pl} \sim 10^{19}$ GeV. Extending 
the range of $c_i$ and $\Lambda$ requires addressing several issues which are 
fully explored in the first part of the paper. In particular, 
 we show that for
 large enough
   positive values of $c$ the vacuum stability analysis of 
 Ref\cite{Casas,DYZ} has to be modified. 
 We also show that, for $c <0$, even close
   to 
   $M_{Pl}$ there can be a significant difference between the higgs mass
   bound 
   obtained in the standard model and the higgs mass bound obtained
  with the presence of higher 
dimensional
   operators. 
   However, for $c>0$ the higgs mass bound in the effective theory differs from
   the standard model value by only a few GeV for high $\Lambda$ even for 
   $c=10$.
  
   Depending on the size and sign of the coupling of the higher 
   dimensional operator, vacuum stability analysis
 gives a band instead of a single value for
 the lower bound on the higgs mass for fixed $\Lambda$.
   In
   general, 
   higgs mass bound from vacuum stability complement the bounds obtained
   from 
   precision electroweak observables. For instance, electroweak precision 
   measurements can be used to obtain an upper bound on the higgs mass for a
   given 
   $\Lambda$ or alternately for a given higgs mass one can obtain a lower
   bound on 
   the scale $\Lambda$. Vacuum stability analysis provide a lower bound 
   on the higgs mass for a given $\Lambda$ and alternately for a given higgs
   mass 
   one can obtain an upper bound on the scale of new physics $\Lambda$.
 
Next we demonstrate that higher dimensional operators that contribute to 
the effective potential can naturally arise in extra dimension theories 
and then 
we study the implications of vacuum stability analysis for 
such theories.

   Analyses of the higher dimension 
   operators in Eq. (1) 
   have been performed by 
   many authors in the literature\cite{bu}. 
   The operator, up to dimension 6, 
   relevant for deriving the lower bound on the higgs mass from vacuum 
   stability\cite{DYZ} is given by 
   \begin{eqnarray} 
   {\cal L}^{new} & = & \frac{c}{\Lambda^2}{( \Phi^+ \Phi - \frac{v^2}{2} )}^3 
   . \ 
   \end{eqnarray} 

   In the presence 
   of the higher dimensional operator in Eq.(2) the 
   tree level Higgs potential can 
   now be written as \cite{zhang} 
   \begin{eqnarray} 
   V_{tree} & = & -\frac{m^{2}}{2}\phi^{2} +\frac{1}{4}\lambda \phi^{4} + 
   \frac{1}{8}\frac{c }{\Lambda^{2}} 
   {(\phi^2 -v^2)}^3, \ 
   \end{eqnarray} 
   which is corrected by the one-loop term, $V_{1loop}$, 
   \begin{eqnarray} 
   V_{1loop}(\mu) & = & \sum_{i} \frac{n_i}{64 \pi^{2}}M_i^4(\phi) 
   \left[\log{\frac{M_i^2(\phi)}{\mu^2}} 
   -C_i\right], \ 
   \end{eqnarray} 
   where 
   \begin{eqnarray} 
   M_i^2(\phi) & = & k_i\phi^2 -k_i' \nonumber.\\ 
   \end{eqnarray} 
   The summation goes over the gauge bosons, the 
   fermions and the scalars of the standard model. 
The values of the 
   constants $n_i$, $k_i$, $k_i'$ 
   and $C_i$ can be found in Refs\cite{Sher,Casas}. 
The full effective potential up to one-loop correction
  is 
   $$ V = V_{tree} + V_{1loop}. $$ 

   Note that the effect of a positive 
   $c $ is to delay the onset of vacuum instability compared to 
   the standard model while the effect of a negative 
   $c $ is to accelerate the onset of vacuum instability. 

   To obtain a lower bound on the higgs boson mass, 
   in the absence of higher dimensional operators, one 
   can take the location of vacuum instability to be as large as $\Lambda$. 
   However, in our approach, for the low energy theory 
   to make sense{\footnote {Effective theory in general will not be valid 
   in the region close to the cutoff scale. One of the examples is the 
   chiral lagrangian of pions where the predictions for processes such as $\pi 
   \pi$ scattering can only be reliable for small momentum transfer relative 
   to the cutoff $\Lambda \sim 4 \pi f_\pi \sim 1$ GeV.}, we should require
   $\phi < 
   \Lambda$. We 
   take the scale of vacuum instability, $\Lambda'$, to be $0.5\Lambda$, so 
   the corrections from operators of dimension greater than six to our 
   result is suppressed by a factor of 
   $\frac{{\Lambda'}^2}{{\Lambda}^2} =0.25$. 

   Since we are dealing with values of the field 
   $\phi$ larger than $v$, we need to consider a renormalization group 
   improved potential 
   for our analysis \cite{Sher,Casas,L,LS,S,Al}. 
   Working 
   with the one-loop effective potential, we 
   consider two-loop 
   running for $\lambda$, the top Yukawa coupling ($g_Y$), gauge couplings 
   and the higgs mass. This procedure re sums all next-to-leading 
   logarithm contributions \cite {Kas}. 
   The various $\beta$ functions to two-loop order 
   can be found in Ref 
   \cite{ME}.

   \begin{figure}[htb] 
   \centerline{\epsfysize 4.2 truein \epsfbox{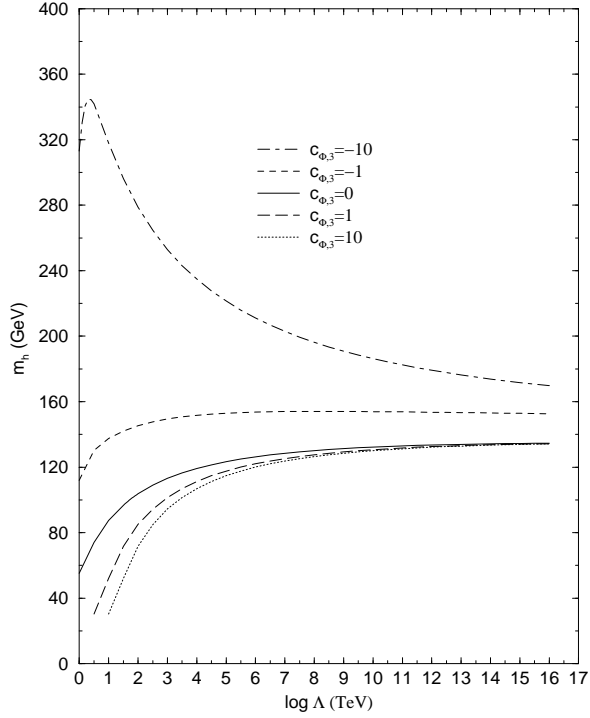}} 
   \caption{The Higgs mass versus the scale of new physics $\Lambda$ } 
   \end{figure} 

   After obtaining the running higgs boson mass, the physical pole mass 
   can be calculated. 
   The relevant 
   equation relating the running mass to the pole mass 
   can be found in Ref \cite{Casas}. 
   The boundary conditions for the gauge couplings and the top quark 
   Yukawa couplings are known at the electroweak scale in terms of the 
   measured values, taking into account 
   the connection between the running top mass 
   and the pole top mass measured at 
   $174$ GeV.

  In the presence of higher dimensional operators, with the scale of 
 vacuum stability
 $\Lambda'$,
     the vacuum stability requirement
\ber 
   V( \Lambda') & = & V(v) \ 
   \eer 
  provides the boundary condition for $\lambda$ at the scale 
   $\Lambda'$, which is given by 
   \begin{eqnarray} 
   \lambda_{eff}(\Lambda') & \approx &-\sum_{i} \frac{n_i}{16 
   \pi^{2}}{k_i}^2(\log{k_i}-C_i) -\frac{1}{2}\frac{\Lambda'^2}{\Lambda^2} 
   c + \frac{2m^2}{\Lambda'^2} , 
   \end{eqnarray} 
   where 
   \begin{eqnarray} 
   \lambda_{eff}(\Lambda') & = & \lambda(\Lambda') 
   -\frac{3}{2} c \frac{v^2}{\Lambda^2} . \ 
   \end{eqnarray} 
One can then run down the higgs coupling $\lambda$ to obtain the higgs mass that ensures
vacuum stability to the scale $\Lambda'$. In general the effective potential 
 increases with $ \phi$ for $\phi >v$ and attains 
a local maximum beyond which it turns around and becomes unstable at the scale $\Lambda'$
where the depth of the potential is the same as for $\phi=v$.

For $c>0$ the higher dimensional operator tends to stabilize the vacuum
 and for a given  higgs mass the onset of vacuum stability can be delayed and
 hence the scale of vacuum
stability can be raised compared to the standard model value. For large enough
values of $c$ the effect of the higher dimensional operator can compensate for
the tendency of the standard model higgs 
potential to become unstable and the instability of the effective potential
disappears for 
 all values of $ v \le \phi \le \Lambda'$. This effect can be demonstrated
in a toy model of new physics  where a scalar field of mass $M$ is added 
to the standard model \cite{Hung}. It was shown in Ref \cite{Hung} that, for 
a given choice of parameters in the effective potential, there is 
a critical value
for the scalar mass ,$M$, below which the vacuum instability disappears.

 In such cases 
the boundary condition
for $\lambda$ at the scale 
 $\Lambda'$, is no longer given by Eq. (6) and one has to numerically 
search for the
minimum higgs mass
 that ensures
   \ber 
   V( \phi) & \ge & V(v) \ 
   \eer 
   for all $\phi \le \Lambda'$. 
In fact if one starts from the boundary condition of Eq. (6) the 
vacuum becomes 
unstable much before $\Lambda'$ and the potential attains a second 
local minimum which is 
deeper than the minimum at $\phi=v$.   

   In Fig. 1 we show the renormalization 
   group improved lower bound on the higgs mass  versus $\Lambda$. For 
   $ c $ we have chosen the values 
   $c = \pm 1$ and $ c = \pm 10$ along with $c=0$ which corresponds 
   to the standard model. In the standard model the
 higgs self coupling
   $\lambda$ 
   increases as we run down from the scale $\Lambda$ to lower scales
  and the lower bound on the
   higgs mass 
   from vacuum stability increases with $\Lambda$. For $c<0$
   the higher dimensional operator tends to destabilize the vacuum and so
   for a given $\Lambda$ the lower bound on the higgs mass for vacuum stability
   is larger than the corresponding standard model value. For
   $c=-1$ 
   the higgs self coupling $\lambda$ continues to increase as we run down
   from 
   $\Lambda$ to lower scales and  the lower bound on the higgs mass 
   increases with $\Lambda$ till $\Lambda \sim 10^8$ TeV, beyond which 
 it decreases slightly with increasing $\Lambda$.
   For $c=-10$ 
   one obtains $\lambda(\Lambda')>1 $ from the boundary condition given  
in Eq. (7). The 
 running of the higgs self coupling in this case is 
   dominated by terms that depend on $\lambda$ and
 and the higgs self coupling decreases as one goes from a
   higher 
   scale to a lower scale. The lower bound on the 
higgs mass, in this case, decreases with
 increasing $\Lambda$ for most values of $\Lambda$. For $c>0$
   the higher dimensional operator tends to stabilize the vacuum and so
   for a given $\Lambda$ the lower bound on the higgs mass for vacuum stability
   is smaller than the corresponding standard model value. In this case also
 the lower bound on
the higgs mass increases with $\Lambda$ as in the standard model. 

An important observation from  Fig. 1 is that 
    the lower bound on the higgs mass 
    in the presence of the higher
 dimensional operator, with $c <0$, can be 
quite different from the standard model value
even for large $\Lambda$ as high as $M_{Pl}$. 
   For $ c >0$, the lower 
   bound on the higgs mass differs from the standard model 
   value by at most a few GeV for high $\Lambda$ though it differs 
significantly 
   from the 
   standard model value for low $\Lambda$. One can understand the behavior
at large $\Lambda$ from the following consideration.
For large $\Lambda$ the effect of the
higher dimensional operator begins to become
 significant beyond some large value of $\phi=\Lambda_1$. If  
 $m_{H1}$ is the higgs mass that ensures vacuum stability to 
the scale $\Lambda_1$ in the standard model then one would expect
the effective potential, with the higher dimensional operator included,
to be stable up to $\phi \sim \Lambda_1$ with the higgs mass $m_{H1}$. If 
$c>0$ the higher dimensional operator stabilizes the vacuum and the vacuum
continues to be stable beyond $\phi=\Lambda_1$ and one can 
obtain vacuum stability to the scale $\Lambda'$ with the higgs mass
$m_{H1}$. However the study of the standard model curve in Fig.1 
shows that the lower bound on the higgs mass increases by smaller
amounts 
as we go to higher values of $\Lambda$. Consequently the difference in the 
higgs mass, $m_{H1}$, that ensures vacuum stability to the scale $\Lambda'$
in the presence of the higher dimensional 
operator and the corresponding standard 
model value is small.
Note that the above argument does not work for $c<0$ because beyond
$\phi \sim \Lambda_1$ the vacuum rapidly becomes unstable and one has to
significantly increase the higgs mass from $m_{H1}$ to ensure that the vacuum
remains stable beyond $\phi= \Lambda_1$ all the way to $\phi=\Lambda'$.

In general, theories with extra dimensions can produce  higher dimensional operators that
contribute to the effective higgs potential considered above. Even though 
the underlying theory describing the physics of extra dimensions is unknown, 
one can construct models where an effective lagrangian describing the coupling of gravity with the standard model fields may be written down. To  
demonstrate the existence of higher dimensional operators that contribute to the effective higgs potential we consider a simple model where only gravity
  is allowed to live in 
$n$ `large' extra dimensions, while the Standard Model(SM) 
fields are confined on a 3-D surface or brane. Gravity then becomes strong 
in the full $4+n$-dimensional space at a scale $M_s$ which  may be 
far below the Planck scale, $M_{Pl}\sim 10^{19}$ GeV. The scales 
$M_s$ and $M_{Pl}$ are related via Gauss's Law:
\begin{equation}
M_{Pl}^2=V_nM_s^{n+2}\,
\end{equation}
with $V_n$ being the volume of the compactified 
extra dimensions. For  the simplest case, termed a symmetric 
compactification, $V_n\sim R^n$. If $M_s$ is in the TeV range then only
$n \ge 2$ is allowed. In such a 
scenario the effects 
of large extra dimensions arise from the interactions 
involving the Kaluza-Klein(KK) excitations of the graviton from 
compactification. At low energies one can construct an effective 
theory of KK gravitons interacting with the standard model fields 
\cite{Han,Guidice}.

Note that we have not included supersymmetry in our framework. Here we are
essentially following the approach of Ref \cite{Nima,Hall} where 
supersymmetry is not necessary from the low energy point of view for 
stabilizing the hierarchy between the weak scale and the Planck scale. 
However supersymmetry may be crucial for the 
self-consistency of the underlying theory 
of extra dimension which could well be a superstring theory. Here we are 
assuming that supersymmetry is broken at some high enough scale above $M_s$ 
and is therefore irrelevant for our analysis.

The contribution to the effective higgs potential in the one loop approximation can be obtained by 
calculating the KK loop attached to 
external higgs legs 
with zero momenta. The loop diagrams are generated from the fundamental
four-point KK-KK-$\Phi\Phi$ (seagull) vertex.  
The loop diagram with six external legs would 
therefore give rise to a higher dimensional operator $ \sim \phi^6$.
The
four-point KK-KK-$\Phi\Phi$ (seagull) 
vertex is given by \cite{Han}   
\begin{equation}
i{\kappa^2\over4}\delta_{ij}
\biggl(C_{\mu\nu,\rho\sigma} m^2_\phi
+C_{\mu\nu,\rho\sigma|\lambda\eta} k_1^\lambda k_2^\eta\biggr)\ ,
\end{equation}
where $k_1, k_2$ are four-momentum of the scalars, $m_{\phi}$ 
is the scalar mass,  $C_{\mu\nu,\rho\sigma}$
is defined in Ref \cite{Han} and 
\begin{equation}
C_{\mu\nu,\rho\sigma|\lambda\eta}
\ =\ {1\over2}\biggl[\eta_{\mu\lambda} C_{\rho\sigma,\nu\eta}
+\eta_{\sigma\lambda} C_{\mu\nu,\rho\eta}
+\eta_{\rho\lambda} C_{\mu\nu,\sigma\eta}
+\eta_{\nu\lambda} C_{\mu\eta,\rho\sigma}
-\eta_{\lambda\eta} C_{\mu\nu,\rho\sigma}
+(\lambda\leftrightarrow\eta)\biggr]\ ,
\end{equation}
with
\be
\kappa^2R^n= 16 \pi (4 \pi)^{n/2}\Gamma(n/2) M_s^{-(n+2)} . \\
\ee
The expression for 
the massive spin 2 KK propagator can be found in Ref \cite{Han,Guidice}.

One can now calculate the contribution to the $\phi^6$ term in the effective 
potential \ber
V_{HD} & =&  \biggl[\frac{m^6_{\phi}\kappa^6 M_s^2}{16 \pi^2} \int^{\infty}_{0}
\sum_{\vec{n}}\frac{B(k)}{(k^2+m_{\vec{n}}^2)^3} k^2dk^2 \biggr]
\frac{\phi^6}{M_s^2}\nonumber\\
B(k) & =&  \frac{k^{12}}{216m_{\vec{n}}^{12}} + \frac{5k^{10}}
{72m_{\vec{n}}^{10}}
+\frac{13k^8}{36m_{\vec{n}}^8} + \frac{263k^6}{216m_{\vec{n}}^6}\nonumber\\
&+&\frac{19k^4}{9m_{\vec{n}}^4} +\frac{61k^2}{36m_{\vec{n}}^2} 
+\frac{61}{54} \
\eer
where we have performed a Wick rotation and
\begin{equation}
m_{\vec{n}^2} = {4\pi^2 \vec{n}^2\over R^2} . \\
\end{equation}
To evaluate the above integral we need to sum over the tower of KK states.
Following Ref\cite{Han} we make the replacement
\be
\sum_{\vec{n}} \to \int^{\infty}_0 \rho(m_{\vec{n}})dm_{\vec{n}}^2 , \\
\ee
where the KK state density as a function of $m_{\vec{n}}$ is given by
\begin{equation}
\rho(m_{\vec{n}})={R^n\  m_{\vec{n}}^{n-2} \over (4\pi)^{n/2}\  \Gamma(n/2)}.
\label{sph}
\end{equation}

The expression for $V_{HD}$ above is both ultraviolet and infrared divergent.
To evaluate $V_{HD}$ we introduce the ultraviolet cut-off $\Lambda \sim M_s$
and the infrared cut-off $ \mu \sim 1/R$. One then obtains
\ber
V_{HD} & = & (16 \pi)^3 \frac{{(4 \pi)^{n}}{\Gamma(n/2)}^2}{16 \pi^2}
\biggl[ \int^1_{0}xdx \int^{1}_{y_{min}}y^{n-2}N(x,y)dy\biggr]
\frac{m_{\phi}^6}{M_s^6}\frac{M_s^4}{M_{Pl}^4}\frac{\phi^6}{M_s^2}\nonumber\\
N(x,y) & = & \frac{x^3}{216y^6} + \frac{5x^2}{72y^5} + \frac{13x}{36y^4} 
+ \frac{263}{216y^3}\nonumber\\
& +& \biggl[- \frac{x^5}{72y^5} - \frac{2x^4}{9y^4} -
 \frac{35x^3}{27y^3} - \frac{97x^2}{36y^2} - \frac{167x}{72y} - 
\frac{19}{216}\biggr]/(x+y)^3 \
\eer
where $y_{min}= M_s^2/M_{Pl}^2$.
The expression above is dominated by the lowest lying KK excitations and one 
ends up with an unreasonably large coefficient for $\phi^6$ if $M_s$ is in the 
TeV range. In fact 
for $n=2$ the leading contribution is
\ber
V_{HD} & = & \frac{512 \pi^3}{675}\frac{m_{\phi}^6}{M_s^6}
\frac{M_{Pl}^6}{M_{s}^6}
\frac{\phi^6}{M_s^2}\
\eer
We would expect that in  the underlying theory of extra dimensions there 
are mechanisms that suppress the contributions of the lowest lying 
KK excitations and a reasonable value for the coefficient for the 
higher dimensional operators can be obtained. We also note that there are
models where the large hierarchy between the lowest KK excitation 
and $M_s$ is absent\cite{raman}. In the absence of 
the knowledge about the underlying theory of extra dimensions 
we replace the coefficient of the higher dimensional operator by an 
unknown coefficient $c$. However, the above analysis clearly 
demonstrates that higher dimensional operators that contribute to 
the effective higgs potential
can naturally arise in theories with extra dimensions.
 
   In  theories with extra dimensions, phenomenological
   constraints 
   from CP violations, FCNC and electroweak precision measurements give a
   bound 
   on $\Lambda=M_s \sim 10 $TeV \cite{Dine,Hall}.
Astrophysical considerations lead to the conservative constraint 
$\Lambda>110$TeV for $n=2$ \cite{astro}. In fact for realistic values of 
the parameters the bound on $\Lambda$ can be raised to
 $\Lambda > 250-350$TeV.  From 
   Fig .1 we see that
if corrections from higher dimensional operators 
are neglected,
the astrophysical constraints 
put limits on the higgs mass $m_h \ge 103-109$ GeV for $n=2$. Hence the 
discovery of the higgs boson
at LEP II will make models of extra dimensions with $n=2$ barely viable.

With the inclusion of the
higher dimensional operators 
 we obtain a range for the 
lower bound on the higgs
   mass, 
   $m_h \ge $ 52-137 GeV for $ c =$ +1 to -1 and 
   $m_h \ge $ 30-317 GeV for $c = $ +10 to -10. 
Thus the discovery of a light Higgs with mass less than 137 GeV
will rule out models which produce large negative c. 
This will have important implications on model building for extra dimension 
theories and consequently on collider signatures of such theories.

  Direct searches from LEP II give 
    $m_h >95.5$ GeV at 95 \% C.L \cite{G}. For a higgs mass 
   of $m_h=100$ GeV the standard model curve in Fig .1 implies an upper
   bound on 
   the scale of 
   new physics $\Lambda \sim 50 $ TeV. However in the effective theory with 
   $c > 0 $ the upper bound on the scale of new physics, $\Lambda$, 
   could be higher than the standard model value by more than a factor of 
   20-60 for 
   $ c = 1-10$. Hence the discovery of a light higgs with $m_h \sim $ 
100 GeV could still allow models of extra dimension physics with $n=2$ to be 
consistent with collider and astrophysical constraints.

   In summary we have re-examined the lower bound on the higgs mass 
   from vacuum stability using an effective lagrangian with
 higher dimensional operators included. We have shown
   that 
   corrections from higher dimensional operators can effect the lower 
   bound significantly. We demonstrated how higher dimensional operators that
contribute to the effective higgs potential can naturally arise in models of extra dimension theories. We also studied the implication of vacuum stability analysis for extra dimension theories.

\section{Acknowledgments}

We thank C. Hill and R. J. Zhang for discussion. This work was supported
in part by NSF of China ( X. Zhang)  and by the Natural  Sciences and 
Engineering Research  Council
of Canada (A. Datta).


\begin{thebibliography}{References} 

   \bibitem{1}For a recent review, see, {\it e. g.}, 
   H. Neuberger, in {\it{ Proceedings of 
   the XXVI International Conference on High Energy Physics}}, Dallas,
   Texas, 
   1992, edited by J. Sanford, AIP Conf. Proc. No. 272 (AIP, New York,
   1992), Vol. 
   II, p. 1360. 

   \bibitem{2}G. 't Hooft, in {\it Recent Developments in Gauge Theories}, 
   Proceedings of the Cargese Summer Institute, Cargese, France, 1979,
   edited by 
   G. 't Hooft et al., NATO Advanced Study Institute Series B; Physics Vol.
   59 
   (Plenum, New York, 1980).
   
   \bibitem{G}T. Greening, hep-ex/9903013.   

   \bibitem{Lep} LEP working group. Available at 
   www.cern.ch/LEPEWWG/plots/winter99. 

   \bibitem{Hall} L. Hall and C. Kolda, hep-ph/9904236. 

   \bibitem{Barb} R. Barbieri and A. Strumia, hep-ph/9905281. 

   \bibitem{Ravi} R. Shekar Chivukula and Nick Evans hep-ph/9907414. 

   \bibitem{Casas} J.A. Casas, J.R. Espinosa and M. Quiros, Phys. Lett. 
   {\bf B 342}, 171 (1995); J.A. Casas, J.R. Espinosa, M. Quiros and A. 
   Riotto, Nucl. Phys. {\bf B 436}, 3 (1995); J.R. Espinosa and 
   M. Quiros, Phys. Lett. {\bf B 353}, 257 (1995); J.A. Casas, J.R. Espinosa
   and M. Quiros, hep-ph/9603227. 

   \bibitem{DYZ} A. Datta, B-L. Young, X. Zhang. Phys. Lett. { \bf B 
385}, 225
 (1996). 

  \bibitem{Nima} N.Arkani-Hamed, S. Dimopoulos and G. Dvali, 
   Phys. Lett.{ \bf B 429}, 263 (1998) and 
   Phys. Rev. {\bf D 59}, 086004 (1999); I. Antoniadis,  N.Arkani-Hamed, 
   S. Dimopoulos and G. Dvali, 
   Phys. Lett.{ \bf B 436}, 257 (1998);
   N. Arkani-Hamed, S. Dimopoulos and J. March-Russell, 
   { hep-th/9809124.}

   \bibitem{Anton}  I. Antoniadis,  Phys. Lett.{ \bf B 246}, 377 (1990).

   \bibitem{Dine} T. Banks, M. Dine and A. Nelson, hep-th/9903019. 

   
   \bibitem{bu} W. Buchm\"uller and D. Wyler, 
   Nucl.Phys. {\bf B268}, 621 (1986); see also 
   K. Hagiwara, R. Szalapski and D. Zeppenfeld, Phys. Lett. 
   {\bf B318}, 155 (1993). 

   \bibitem{zhang} X. Zhang, Phys. Rev. {\bf D47}, 3065 (1993). 

   \bibitem{Sher} M. Sher, Physics Reports {\bf 179}, 273 (1989). 

   

   \bibitem{L} M. Lindner, Z. Phys. {\bf C 31}, 295 (1986). 

   \bibitem{LS} M. Lindner, M. Sher and H.W. Zaglauer, Phys. Lett. {\bf B 
   228}, 139 (1989). 

   \bibitem{S} M. Sher, Phys. Lett. {\bf B 
   317}, 159 (1993); Addendum: hep-ph/9404347. 


   \bibitem{Al} G. Altarelli and G. Isidori, Phys. Lett. {\bf B 337}, 141 
   (1994). 

   \bibitem{Kas} B. Kastening, Phys. Lett. {\bf B 283}, 287 
   (1992); M. Bando, T. Kugo, N. Maekawa and H. Nakano, Phys. Lett. 
   {\bf B 301}, 83 (1993); Prog. Theor. Phys. 90 (193). 

   \bibitem{ME} C. Ford, D.R.T. Jones, P.W. Stephenson and M.B. Einhorn, 
   Nucl. Phys. {\bf B 395}, 17 (1993). 
 
   \bibitem{Hung} P.Q. Hung and Marc Sher, Phys. Lett. {\bf B 374}, 138 (1996).
   
\bibitem{Han} T. Han, J.D. Lykken and R. Zhang, Phys. Rev. {\bf D59}, 
105006 (1999).

\bibitem{Guidice} G.F. Giudice, R. Rattazzi and J.D. Wells, Nucl. Phys. 
{\bf B 544} 3 1999. 

\bibitem{raman} L. Randall and R. Sundrum, hep-ph/9905221.  

\bibitem{astro} 
S.~Cullen and M.~Perelstein, Phys. Rev. Lett { \bf 83}, 268 1999 ;
L.J.~Hall and D.~Smith, hep-ph/9904267; 
V.~Barger, T.~Han, C.~Kao and R.J.~Zhang, hep-ph/9905474;
G.C. McLaughlin and J.N. Ng, hep-ph/9909558. 
\end{thebibliography}
   \end{document}